\documentclass[11pt,a4paper]{article}
\usepackage[dvips]{graphicx,epsfig,color}
\usepackage{amsmath, amssymb}
\usepackage{bm}

\def\fb0{\mbox{\boldmath $0$}}
\def\b1{\mbox{\boldmath $1$}}

\def\bx{\mbox{\boldmath $x$}}
\def\by{\mbox{\boldmath $y$}}

\def\E{\mbox{E}}

\def\N{\mbox{N}}

\def\I{\mbox{I}}

\def\U{\mbox{U}}

\def\Bi{\mbox{Bi}}

\def\KL{\mbox{KL}}

\def\Max{\mbox{Max}}
\def\Min{\mbox{Min}}

\def\clC{\cal{C}}

\newcommand{\Argmax}{\mathop{\rm Argmax}}

\newcommand{\Epost}{\mathop{\rm E_{post}}}

\newcommand{\tsum}{\mathop{\textstyle{\sum}}}
\newcommand{\tprod}{\mathop{\textstyle{\prod}}}

\begin{document}

\def\baselinestretch{1.45}\large\normalsize
\parindent=2ex

\vspace*{2ex}
\begin{center}
{\Large A Pair of Novel Priors for Improving and Extending}

{\Large the Conditional MLE}

\vspace*{2ex}

${\rm T. Yanagimoto^{1*}}$ and ${\rm Y. Miyata^2}$
\end{center}

\begin{center}
1. {\it Institute of Statistical Mathematics, Tokyo, Japan}

2. {\it Faculty of Economics, Takasaki City University of Economics, Gunma, Japan}

* E-mail address: takemiyanagimoto@gmail.com 
\end{center}

\begin{abstract}
A Bayesian estimator aiming at improving the conditional MLE is proposed by introducing a pair of priors.
After explaining the conditional MLE by the posterior mode under a prior,
we define a promising estimator by the posterior mean under a corresponding prior.
The prior is equivalent to the reference prior in familiar models.
Advantages of the present approach include two different optimality properties of the induced estimator, the ease of various extensions and the possible treatments for a finite sample size.
The existing approaches are discussed and critiqued.
\end{abstract}

\vspace*{1ex}
\noindent
{\bf AMS 2020 subject classification}: Primary 62F10; secondary 62F15 

\noindent
{\bf Keywords}: Conditional MLE, nuisance parameter, optimum predictor, prior elicitation, reference prior, separate inference

\vspace*{5mm}
\noindent
{\bf 1. Introduction}
\setcounter{section}{1}
\setcounter{equation}{0}

The conditional maximum likelihood estimator (MLE) is a standard technique in separate inference.
Let $\{p(\bx|\theta); \theta \in\Theta\}$ be a family of sampling densities of a sample vector of size $n(\ge 2)$.
The parameter is decomposed into $\theta=(\lambda,\psi)$, where both the components will be scalars for simplicity.
Suppose that there exists an ancillary statistic $t(=t(\bx))$ such that the factorization property:
\begin{equation}
p(\bx|\lambda,\psi) \,=\, pm(t|\lambda,\psi)pc(\bx|t,\psi)
\label{FactAssump}
\end{equation}
holds for every $\bx$ and $\theta$.
Then the conditional MLE of $\psi$, $\hat\psi_{CML}$, is defined as ${\Argmax}_\psi pc(\bx|t,\psi)$.
We intend to estimate of both components of $\theta$.
Then an estimator of $\lambda$ is necessary, and we define it by ${\Argmax}_\lambda p(\bx|\lambda, \psi)$.
In familiar models, it holds that $\lambda$ is estimation orthogonal to $\psi$ (Lindley, 1996), that is,
\begin{equation}
\Argmax_\lambda p(\bx|\lambda,\psi) \,=\, \hat\lambda_{ML} \,=\, t(\bx)
\label{EstOrht}
\end{equation}
The estimator of $\lambda$ will be written as $\hat\lambda_{ML}$, when this property holds

Our aim is to construct a Bayesian estimator which is expected to perform better than $(\hat\lambda_{ML},\hat\psi_{CML})$.
For this purpose, we introduce a prior $\pi_m(\lambda,\psi)$ such that the posterior mode becomes $(\hat\lambda_{ML},\hat\psi_{CML})$.
Then we define another prior proportional to  $\pi_m(\lambda,\psi)\pi_J (\lambda,\psi)$, where $\pi_J(\lambda,\psi)$ is the Jeffreys prior.
This form is selected so that the posterior mean is close to  $(\hat\lambda_{ML},\hat\psi_{CML})$.

Our approach is naive and simple, and one may wonder why such an attempt is not seen in the existing literature.
We emphasize here that our approach is largely different from existing ones.
Traditionally, a conditional inferential procedure focuses only on $\psi$, when the factorization property \eqref{FactAssump} holds.
The remaining component $\lambda$ is regarded as a nuisance, and is not estimated.
This treatment makes a comparison study between the conditional MLE and a Bayesian estimator difficult, since the the latter estimates both the components.
Note that $(\hat\lambda_{ML},\hat\psi_{CML})$ is attractive.
In fact, it dominates the MLE of $(\lambda,\psi)$, $(\hat\lambda_{ML},\hat\psi_{ML})$, in familiar models (Yanagimoto and Anraku, 1989).
Existing Bayesian approaches to constructing inferential procedures acceptable for the frequentist are based on a non-informative or an objective prior.
Emphasis has been placed on an appealing prior itself, and least attentions has been paid to a prior implying an attractive estimator.

We will find at least three advantages of the present approach.
One is that the induced estimator satisfies two different optimality properties.
This is to be compared with the fact that a usual Bayesian estimator satisfies one.
The second is the ease of various extensions, which allow us to apply the procedures to a variety of models.
The third is that explicit forms of a pair of priors are obtained for a fixed sample size.

The present paper is organized as follows.
A pair of novel priors are defined for implying the posterior mode and the posterior mean, and their basic properties are presented in Section 2, which is followed by examples of familiar models in Section 3.
Section 4 provides us with various extensions and also with additional examples.
Finally, the conditional MLE and the reference prior are discussed and critiqued based on the proposed priors.

\vspace*{5mm}
\noindent
{\bf 2. Prior elicitation and basic properties }
\setcounter{section}{2}
\setcounter{equation}{0}

A family of sampling densities $\{p(\bx|\theta); \theta \in\Theta\}$ with  $\theta=(\lambda,\psi)$ is assumed to satisfy the factorization property \eqref{FactAssump}.
We begin by defining a pair of priors, and then present basic properties.

\vspace*{5mm}
\noindent
{\bf 2.1. Definitions}

Let $t$ be a suitably chosen ancillary statistic for $\psi$ such that a family of distributions satisfies regularity conditions on the factorization property.
The default choice of such a statistic $t$ is the MLE of $\lambda$.
This default choice is widely applicable in practical examples.
Other possible choices will be discussed in the subsection 4.2.
Write the MLE of $\lambda$ for a given $\psi$ as $\hat\lambda_{ML}=\hat\lambda_{ML}(\psi)$, and also the marginal likelihood of $t$ as $pm(t|\lambda,\psi)$.
The profile marginal likelihood for $\psi$ is expressed as $pm(t|\hat\lambda_{ML}(\psi),\psi)$.
Then we define a prior by the reciprocal of profile marginal likelihood, as follows.

\vspace*{3mm}
\noindent
{\bf Definition 1}. 
A prior function for $(\lambda, \psi)$ is defined in terms of the reciprocal of $pm(t|\hat\lambda_{ML}(\psi),\psi)$ as
\begin{equation}
\pi_m(\lambda, \psi) \,=\,\pi_m(\psi) \,\propto\, \frac{1}{pm(t|\hat\lambda_{ML}(\psi),\psi)}. 
\label{CorePrio}
\end{equation}
This prior will be called the profile marginal likelihood (PML) prior. 

\vspace*{3mm}
When the regularity condition on the estimation orthogonality in \eqref{EstOrht} is satisfied, $\hat\lambda_{ML}(\psi)$ is independent of $\psi$, and will be written as $\hat\lambda_{ML}$.

Note that the necessity of the separate inference comes from the possible excess of the maximized likelihood.
This point will be reviewed in the subsection 2.3.
The reason why the reciprocal form is employed is that the possible excess of the profile marginal likelihood is to be eliminated or to be reduced.

A definition of a stronger notion of an ancillary statistic for $\psi$, an s-ancillary statistic, requires that the profile marginal likelihood of $t$ does not depend on $\psi$, see Cox and Hinkley (1974) and Lindsey (1996).
This condition yields that $\pi_m(\psi)\propto 1$.
We obtain a direct relation between the posterior mode and the conditional MLE.
For this purpose we need a regularity condition.
\begin{equation}
\Max_{(\lambda,\psi)} p(\bx|\lambda,\psi)\pi_m(\psi) \,=\, \Max_\psi \{\Max_\lambda p(\bx|\lambda,\psi)\pi_m(\psi)\} 
\label{regargmax}
\end{equation}

\vspace*{3mm}
\noindent
{\bf Proposition 2.1}.
Write the posterior mode of $(\lambda,\psi)$ under a prior function $\pi_m(\psi)$ as $(\hat\lambda_{Mod},\hat\psi_{Mod})$.
Then under the factorization property in \eqref{FactAssump} and \eqref{regargmax} it holds that 
$$
(\hat\lambda_{Mod},\hat\psi_{Mod}) \,=\, (\hat\lambda_{ML}(\hat\psi_{CML}),\hat\psi_{CML}). $$ 
where $\hat\psi_{CML}=\Argmax_\psi pc(\bx|t,\psi)$ and $\hat\lambda_{ML}(\hat\psi_{CML})=\Argmax_\lambda pm(t|\lambda,\hat\psi_{CML})$ denote the conditional MLE of $\psi$ given $t$ and the MLE of $\lambda$ given $\hat\psi_{CML}$, respectively.

\vspace*{3mm}
When the estimation orthogonality in \eqref{EstOrht} is satisfied, the posterior mode becomes simpler, and is written as $(\hat\lambda_{ML},\hat\psi_{CML}) $.

We emphasize here that the posterior mode provides us with an estimator of $\lambda$ as well as that of $\psi$.
On the other hand, the traditional conditional inference treats $\lambda$ as a nuisance.
This tradition makes researches on the estimators of $\lambda$ and $(\lambda,\psi)$ less familiar.
Further, the conditional MLE of $\psi$ is not a subject to be improved, though such an attempt may be tough in the frequentist framework.
It is an attractive research subject to improve the conditional MLE in the Bayesian framework.

Since the conditional MLE of $\psi$ is explained as a part of the posterior mode of $(\lambda,\psi)$, our aim is realized by showing that the posterior mean is expected to behave better than the posterior mode.
For this purpose it is necessary to choose two different priors under which the posterior mean and the posterior mode are induced.
The Jeffreys prior function $\pi_J(\lambda,\psi)$ for $(\lambda,\psi)$ will be chosen as a default non-informative prior to define a prior matching the posterior mean and the posterior mode.
Possible other choices will be discussed in the subsection 4.3.

\vspace*{3mm}
\noindent
{\bf Definition 2}.
Another prior function derived from the profile marginal likelihood of an ancillary statistic is given by
\begin{equation}
\pi_M(\lambda,\psi) \,\propto\, \pi_m(\psi) \pi_J(\lambda, \psi). 
\label{MRLDef}
\end{equation}
This prior will be called the matching profile marginal likelihood (MPML) prior. 

\vspace*{3mm}
By definition, the MPML prior can depend on the statistic $t$ and the sample size $n$.
If one favors a non-informative, objective one, this dependency may be to be avoided.
Our standpoint is more practical, and we intend to define an estimator behaving favorably.
Note that the conditional inference owes essentially to the statistic $t$.

Write the posterior density induced from the MPML prior as $\pi_M(\lambda,\psi|\bx)$, and choose an estimand $(g_1(\lambda, \psi), g_2(\lambda,\psi))$.
For notational convention, the posterior mean of $g_i(\lambda, \psi)$ will be written as $\Epost[g_i(\lambda, \psi)]$ for $i=1$, 2.

\vspace*{3mm}
\noindent
{\bf Definition 3}. 
The component $g_i(\lambda, \psi)$ for $i=1$, 2 is estimated by the posterior mean under the density $\pi_M(\lambda,\psi|\bx)$
\begin{equation}
\hat g_i(\lambda, \psi) \,=\, \Epost[g_i(\lambda, \psi)].
\label{CorePrio}
\end{equation}

\vspace*{3mm}
The performance of the posterior mean depends on the choice of an estimand.
When the sampling density is in the exponential family, a canonical parameter will be chosen as an estimand.
A reason will be discussed in the subsection 2.3.
We will regard that two estimators are equivalent when the plug-in predictors are equivalent. 

\vspace*{5mm}
\noindent
{\bf 2.2. Asymptotic relations}

To explore an asymptotic relation rigorously in a general setting, we state this problem in a general form in this subsection.
The parameter $\theta$ is in $R^d$ for $d\ge2$ in this subsection.

Let $p(\bx|\theta)$ be a density function satisfying  necessary regularity conditions stated later.
For notational convention, we will write $h(\theta)=-\log p(\bx|\theta)/n$.
Write its partial derivative with respect to $\theta_i$ and $\theta_j$ as $h_{ij}(\theta)$, and the $(i,j)$ element of the inverse of the matrix $\big(h_{ij}(\theta) \big)$ as $h^{ij}(\theta)$, respectively.
Suppose that an estimator $\hat \theta$ satisfies the condition
\begin{equation} 
\hat\theta \,-\, \hat\theta_{ML} \,=\, O(1/n) 
\label{AsympCond1}
\end{equation}
where $\hat\theta_{ML}$ is the MLE. 
Miyata (2004) generalized an asymptotic expansion of the posterior mean of $g(\theta)$ about $g(\hat\theta_{ML})$ into that about  $g(\hat\theta)$  for an arbitrary estimator $\hat\theta$ satisfying the condition \eqref{AsympCond1}.
Regularity conditions are necessary to apply the Laplace approximation, as in Miyata (2004).
We will assume such conditions without explaining them explicitly.
Write the partial derivative $\partial g(\theta)/\partial \theta_i$ simply as $g_i(\theta)$.

\vspace*{3mm}
\noindent
{\bf Lemma 2.1} (Miyata, 2004).

Assume that an estimator $\hat\theta$ satisfies the condition \eqref{AsympCond1}.
The asymptotic expansion of $\Epost [g(\theta)] -g(\hat\theta)$ up to the order $O(1/n)$ is given in terms of the estimator $\hat \theta$ as 
\begin{equation}
\left. \frac{1}{n}\left[\tsum g_i(\theta) h^{i j} (\theta) \left\{ \frac{\pi_j(\theta)} {\pi(\theta)} \,-\, n h_j(\theta)
\,-\, \frac{1}{2} \tsum h^{r s}(\theta)  h_{rsj }(\theta) \right\} \,+\, \frac{1}{2} \tsum h^{ij} (\theta) g_{ij}(\theta)  \right] \right|_{\theta=\hat\theta} 
\label{KassFor}
\end{equation}
where $\pi_i(\theta )$ and $h_{rsj}(\theta)$ denote the partial derivatives of $\pi(\theta)$ and $h(\theta)$ with respect to $\theta_i$ and $\theta_r$, $\theta_s$, and $\theta_j$, respectively. 

The remaining order is $O(1/n^2)$. 
Note that the term $h_j(\theta)$ vanishes when $\hat\theta$ is the MLE. 

\vspace*{3mm}
\noindent
Proof. The symbol $A^{\top}$ denotes the transpose of matrix $A$, and $\mathrm{tr}(A)$ denotes the trace of matrix A. Although we here derive equation (2.4) formally, we can justify this asymptotic expansion under appropriate conditions. 
We let $h^{*}(\theta )=-\log\{  p(\bx|\theta)\pi (\theta) \}$ be minus the log-posterior function. It follows from condition (2.5) that 
\begin{align}
\frac{\partial}{\partial \theta}h^{*}(\hat{\theta})&=\frac{\partial}{\partial \theta}h(\hat{\theta})-\frac{1}{n}\frac{\partial}{\partial \theta}\log \pi (\hat{\theta}) \notag\\
 &=\frac{\partial}{\partial \theta}h(\hat{\theta}_{ML})+O(1/n) \notag \\
 &=O(1/n). \notag
\end{align}
This means that $\hat{\theta}$ is an asymptotic mode for $h^{*}$ of order $n^{-1}$. Because $(\partial^{2}/\partial \theta \partial \theta^{\top})h^{*}(\hat{\theta} )=(\partial^{2}/\partial \theta \partial \theta^{\top})h(\hat{\theta} ) +O(1/n)$ from condition (2.5), using Theorem 5 of Miyata (2004), we have
\begin{align*}
\Epost[g(\theta)]-g(\hat{\theta})=&-\frac{\partial}{\partial \theta}g(\hat{\theta})^{\top}\left[ \frac{\partial^{2}}{\partial \theta \partial \theta^{\top}}h^{*}(\hat{\theta}) \right]^{-1}\frac{\partial}{\partial \theta}h^{*}(\hat{\theta}) \\
  &-\frac{1}{2n}\tsum h_{ijk}^{*}(\hat{\theta})h^{*iq}(\hat{\theta})h^{*jk}(\hat{\theta}) \frac{\partial}{\partial \theta_{q}}g(\hat{\theta}) \\
 &+\frac{1}{2n}\mathrm{tr}\left( \frac{\partial^{2}}{\partial \theta \partial \theta^{\top}}g(\hat{\theta})  \left[ \frac{\partial^{2}}{\partial \theta \partial \theta^{\top}}h^{*}(\hat{\theta}) \right]^{-1}\right) +O(1/n^{2}) \\
=&-\tsum g_i(\theta) h^{i j} (\theta) \frac{\partial}{\partial \theta_j}\left( h(\theta )-\frac{1}{n}\log\pi (\theta ) \right) \biggr|_{\theta =\hat{\theta}} \\
&-\frac{1}{2n}\tsum g_{i} (\theta)h^{ij} (\theta)h^{rs} (\theta)h_{rsj} (\theta) \biggr|_{\theta =\hat{\theta}}+\frac{1}{2n}\tsum h^{ij}(\theta)g_{ij}(\theta) \biggr|_{\theta =\hat{\theta}} \\
&+O(1/n^{2}), 
\end{align*}
which completes the proof. 
\vspace*{3mm}

For notational convention, the form \eqref{KassFor} will be abbreviated as 
$$
(1/n)[\tsum g_i(\theta) h^{i j} (\theta) \{(\partial/\partial \theta_j) ( \log \pi(\theta)\,-\, n h(\theta)) \} \,+\,S(g(\theta),h(\theta)) ] |_{\theta=\hat\theta} $$
or more simply as $T(\pi (\theta),g(\theta),h(\theta))|_{\theta=\hat\theta} $.

Consider two prior functions, $\pi_r(\theta)$ and  $\pi_N(\theta)$ satisfying regularity conditions. 
We define  an estimator $\hat\theta_r= \Argmax \{p(\bx|\theta)\pi_r(\theta) \}$ by the posterior mode under the assumption of a prior function $\pi_r(\theta)$.
Note that this estimator is the maximum penalized likelihood estimator or the posterior mode, when $\pi_r(\theta)$ is regarded as a penalty term or a prior function, respectively.

Let $\pi_N(\theta)$ be a suitably chosen non-informative prior, and set $\pi_A(\theta) =\pi_r(\theta) \pi_N(\theta)$.
Then consider the two posterior means under the assumption of two prior functions, $\pi_N(\theta)$ and $\pi_A(\theta)$.
To specify a prior function we will write the posterior mean under $\pi_A(\theta|\bx)\propto p(\bx|\theta)\cdot \pi_A(\theta)$ as $\E_{\rm post}^A[g(\theta)]$

\vspace*{3mm}
\noindent
{\bf Theorm 2.1}.

Let $g(\theta)$ be a smooth function of $\theta$.
Suppose that the posterior mode $\hat\theta_r$ satisfies the condition \eqref{AsympCond1}.
Then it follows that
\begin{equation}
\E_{\rm post}^A[g(\theta)] \,-\, g(\hat\theta_r)\,=\,
\E_{\rm post}^N[g(\theta)] \,-\, g(\hat\theta_{ML})  \,+\, O(1/n^2). 
\label{MiyataCorE}
\end{equation}

\vspace*{3mm}
\noindent
Proof.  
The definition of $\hat\theta_r$ implies
$$ 
-n  \partial h (\theta)/ \partial \theta \,+\, 
  \partial \log \pi_r (\theta)/ \partial \theta \,=\, 0 $$
which yields for every $j$ that
$$
\left. \frac{\partial}{\partial \theta_j} \{ \log \pi_A(\theta) \,-\, n h(\theta) \} \right|_{\theta=\hat\theta_r}
\,=\, \left. \frac{\partial}{\partial \theta_j} \log \pi_N(\theta) \right|_{\theta=\hat\theta_r} . $$
Thus the difference $\E_{\rm post}^A[g(\theta)] \,-\, g(\hat\theta_r) \,-\, \{ \E_{\rm post}^N[g(\theta)] \,-\, g(\hat\theta_{ML}) \} $ is expressed as 
$$
T(\pi_A (\theta),g(\theta),h(\theta))|_{\theta=\hat\theta_r}  \,-\, T(\pi_N (\theta),g(\theta),h(\theta))|_{\theta=\hat\theta_{ML}}  \,=\, O(1/n^2)$$
This completes the proof.

\vspace*{3mm}
Write the order of the asymptotic equivalence between $\E_{\rm post}^N[g(\theta)]$ and $g(\hat\theta_{ML})$ as $O(n^{-\alpha})$.
Then the power $\alpha$ takes often one of the three values, 1, 3/2 and 2.
The assumption of Lemma 2.1 requires that the asymptotic order between $g(\hat\theta_r)$ and $g(\hat\theta_{ML})$, is greater than or equal to 1. 
The following corollary is a direct consequence of \eqref{MiyataCorE} in Theorem 2.1.

\vspace*{3mm}
\noindent
{\bf Corollary 2.1.}

Assume that the regularity conditions necessary in Lemma 2.1, and suppose that the asymptotic equivalence order between $\E_{\rm post}^N[g(\theta)]$ and $g(\hat\theta_{ML})$ is $O(n^{-\alpha})$ for some $0< \alpha \leq 2$.
Then that between $\E_{\rm post}^A[g(\theta)]$ and $g(\hat\theta_r)$ is also  $O(n^{-\alpha})$.

\vspace*{3mm}
\vspace*{3mm}
An important case of $O(n^{-2})$ appears, when the sampling density $p(\bx|\theta)$ is in the exponential family and $\pi_N(\theta)$ is the Jeffreys prior function, 
This case will be reviewed in Example 3.2.
Cases of $O(n^{-3/2})$ were discussed extensively in Ghosh and Liu (2011).

\vspace*{5mm}
\noindent
{\bf 2.3. Role of the posterior mean}

The proposed estimator is the posterior mean under the MPML prior function $\pi_M(\lambda,\psi)$.
It is expected to perform favorably compared with other existing estimators such as the MLE, the conditional MLE and the posterior mean under the assumption of the Jeffreys prior.
Since the conditional MLE is closely related to the posterior mode under the assumption of $\pi_m(\psi)$ in the familiar models, this posterior mode is also of our interest. 

A standard Bayesian estimator is given by the posterior mean of a suitably chosen estimand $(g_1(\lambda,\psi), g_2(\lambda,\psi) ) \in R^2$.
A convenient property is obtained by choosing a canonical parameter $(\xi,\psi)$ as an estimand in the exponential family.
Let $(\hat\lambda,\hat\psi)$ be the estimator of $(\lambda,\psi)$  corresponding to $(\hat\xi,\hat\psi)$. 
Then the plug-in predictor $p(\by|\hat\lambda,\hat\psi)$ satisfies an optimality property.
To explain this property, write an arbitrary predictor given $\bx$ as $p(\by|\bx)$, and set the Kullback-Leibler divergence between $p(\by|\bx)$ and $p(\by|\lambda,\psi)$ as $\KL(p(\by|\bx),p(\by|\lambda,\psi))$.
Then the plug-in predictor of the proposed estimator minimizes the Bayesian risk under the Kullback-Leibler loss 
$\Epost[\KL(p(\by|\bx),p(\by|\lambda,\psi))]$ among all the predictors $p(\by|\bx)$, see Corcuera and Giummole (1999) and Yanagimoto and Ohnishi (2009).

No optimality property of the plug-in predictor of the MLE is known.
More seriously, the least favorable property of the plug-in predictor of the MLE was pointed out in Yanagimoto and Ohnishi (2011).
Let $p(\bx|\theta)$ be a general sampling density in the exponential family, and consider the class of predictors satisfying the saddlepoint property
$$
{\clC} \,=\, \{p(\by|\bx) \,|\, \Epost[\KL\big(p(\by|\bx),p(\by|\theta) \big) \,-\, \log\{p(\bx|\bx)/p(\bx|\theta) \}] \,=\, 0 \} $$
Note that the plug-in predictor of the MLE is in this class of predictors, since it is driven from the well-known sample-wise identity $\KL\big(p(\by|\hat\theta_{ML}),p(\by|\theta)\big) = \log\{p(\bx|\hat\theta_{ML})/p(\bx|\theta)\}$
due to Kullback (1959).
This sample-wise equality yields
\begin{equation}
p(\by|\hat\theta_{ML}) \,=\, \Max_{p(\by|\bx) \in {\clC} }\Epost[\KL(p(\by|\bx),p(\by|\theta)]
\label{WsttMLE}
\end{equation}
while the optimality property of the posterior mean of the canonical parameter $\theta$ yields
\begin{equation}
p(\by|\hat\theta) \,=\, \Min_{p(\by|\bx) \in {\clC} }\Epost[\KL(p(\by|\bx),p(\by|\theta)].
\label{BstPosm}
\end{equation}
The former equality indicates the need for the careful use of the MLE; the maximization of the likelihood is directly associated with that of the Kullback-Leibler divergence, when the sampling density is in the exponential family.
On the other hand, the latter equality \eqref{BstPosm} shows the promising role of the posterior mean of the canonical parameter.
Note, however, that detailed case by case studies on the risk comparison will be required to reach a definite conclusion.

The criticism against the MLE is expected to be applied to that against the posterior mode.
In fact, the MLE is the same as the posterior mode under the assumption of the uniform prior.
It is known that the posterior mode satisfies an optimality under the zero-one loss, which corresponds with the optimality property of the posterior mean under the squared loss, see Robert (2001) for example.

It should be emphasized, however, that the posterior mean is associated with the optimum predictor, as reviewed above.
On the other hand, any optimality property of the posterior mode under a loss relating to a predictor is not seen in the literature.
In light of the recent interest in the predictor, this optimality property is important.
It is expected that the behaviors of the posterior mean under the frequentist criteria are promising

\vspace*{5mm}
\noindent
{\bf 2.4. Choice of an estimand}

We remarked in the previous subsection that the importance of the suitable choice of an estimand, and learned the choice of the canonical parameter in the exponential family as an estimand leads us to the optimum plug-in predictor.
This indicates the need for the careful choice of an estimand, when the posterior mean is used as an estimator.

When an informative or a non-informative prior is elicited, it is wise to choose an explicit form of a parameter so that each component is easily understood.
Further, a component of a parameter is hoped to be orthogonal to the other component.
The location-scale parameter and the location-dispersion parameter are typical examples of the naive parametrization.

When we intend to derive a favorable estimator, a suitable choice of an estimand becomes important.
Such a choice will depend on a selected loss.
Losses induced from the plug-in predictors provide us with candidates of an estimand.
We suggested the Kullback-Leibler loss as a reasonable one.
An alternative choice of an estimand may be the expectation parameter in the exponential family.
It satisfies another optimality property under the dual version of the Kullbacj-Leibler loss, though the optimality property holds only among plug-in predictors.
It looks that the expectation parameter in the exponential family  is not of a simple form even in the normal distribution.

When a suitable choice of an estimand is unavailable, a practical method for yielding an estimator is the posterior mode under the assumption of a prior function $\pi_m(\psi)$ in \eqref{AsympCond1}.

\vspace*{5mm}
\noindent
{\bf 3. Examples}
\setcounter{section}{3}
\setcounter{equation}{0}

To explain the possible usefulness of the novel MPML prior, we examine its explicit forms in selected families of distributions, which will be followed by its extensions in the following section.
 
\vspace*{3mm}
\noindent
{\bf Example 3.1} (Normal distribution).
As is usual, the example of the normal distribution provides us with a simple, illustrative one of a newly introduced procedure.
Let $\bx$ be a sample vector of size $n$ from the normal distribution $\N(\lambda,1/\psi)$, and set an ancillary statistic for $\psi$ as $t=\bar x$.
Then the prior functions $\pi_m(\psi)$ and $\pi_M(\lambda,\psi)$ become $1/\sqrt{\psi}$ and $1/\psi$, respectively.
The posterior mode of $(\lambda, \psi)$ under $\pi_m(\psi)$ is $(\hat\lambda_{Mod}, \hat\psi_{Mod})=(\bar x, 1/s^2)$ with $s^2$ being the unbiased sample variance.
The posterior mean of the canonical parameter $(\psi\lambda,\psi)$ under $\pi_M(\lambda,\psi)$ is $\E_{\rm post}^M [(\psi\lambda,\psi)]= (\bar x/s^2,1/s^2)$, which is equivalent to the above posterior mode.
This estimator satisfies the optimum property, as was stated in the subsection 2.3.
It dominates the MLE under the frequentist losses, see Yanagimoto and Anraku(1989) for example.

These simple calculations lead us to various implications.
First, $\pi_M(\lambda,\psi)$ is equivalent to the reference prior function introduced by Bernald (1979).
Note that an explicit form os the reference prior depends on the choice of the component of interest.
We will treat only the case where $\psi$ is of interest.
Second, the estimator $\hat\psi$ is equal to the conditional MLE of $\psi$ given $\bar x$.

Another interesting fact is that the reciprocal of the estimator of $\psi$, $1/\hat\psi$, is an unbiased estimator of $1/\psi$, though the posterior mean of $\psi$ instead of $1/\psi$ is taken.
Such a unbiasedness property is observed in selected families of distributions including the gamma distribution.

\vspace*{3mm}
\noindent
{\bf Example 3.2} (Exponential dispersion model).
Let $\lambda$ be the mean of an observation, and $c(\lambda)$ be a canonical link function of $\lambda$.
The density function in the exponential dispersion model (Jorgensen, 1997) is expressed as 
\begin{equation}
 p(\bx|\lambda,\psi) \,=\,\exp \left[ \psi \tsum \{x_ic(\lambda) \,-\, M(c(\lambda) ) \,-\, N(x_i) \}\right] \tprod a(x_i) \exp n k(\psi) 
\label{ExpDisp}
\end{equation}
where $M(\eta)$ with $\eta=c(\lambda)$ is a convex function of $\eta$, $N(x)= \sup_\eta \{ x \eta - M(\eta) \}$ is the conjugate convex function to $M(\eta)$. 
It is a subfamily of the exponential family with the sufficient statistic $(\bar x, \tsum N(x_i))$ and the canonical parameter $(\psi c(\lambda), \psi)$.
This model contains the three important families of distributions, the normal, the inverse Gaussian and the gamma distributions, and allows us to treat them in a unified way.
Explicit forms of the canonical link function $c(\lambda)$ are expressed as $\lambda$, $-1/\lambda$ and $-/2\lambda^2$ in the normal, the inverse Gaussian and the gamma distributions, respectively.

Note that the mean of $x$ is written by $\lambda=M'(\eta)$.
The density function in \eqref{ExpDisp} can be factored into
\begin{eqnarray}
\exp n \psi \{\bar x c(\lambda) \,-\, M(c(\lambda) &-& N(\bar x) \} a(\bar x) \exp k(n \psi) 
\nonumber \\
 \cdot \exp \psi \{n N(\bar x) & -& \tsum N(x_i) \} \frac{\prod a(x_i)}{  a(\bar x)} \exp \{ n k(\psi)- k(n \psi) \} 
\label{ExpDFact}
\end{eqnarray}
which will be written as $pm(\bar x|\lambda, \psi) \cdot pc(\bx|\bar x, \psi)$.
The component $\lambda$ is estimation orthogonal to $\psi$ in \eqref{EstOrht}, and  the MLE of $\lambda$ is $\bar x$.
It follows that $\pi_m(\psi) \propto \exp -k(n\psi)$.
The Fisher information matrix is diagonal, whose diagonal elements are $\I_{11}= n\psi c'(\lambda)$ and $\I_{22}= nk''(\psi)$.
These facts imply the MPML prior function as
$$
\pi_M(\lambda, \psi)\propto \sqrt{c'(\lambda)k''(\psi)}\exp -k(n\psi). $$

When the sampling density follows the inverse Gaussian distribution, it follows that $k(n\psi)=(1/2)\log(n\psi)$, and also that the MPML prior function $\pi_M(\lambda.\psi)$ is proportional to $\lambda^{-3/2}\psi^{-1/2}$.
Consequently, it is equivalent to the reference prior function,

When the sampling density follows the gamma distribution, $k(n\psi)$ is expressed as $-\log \Gamma(n\psi) +n\psi\log(n\psi)-n\psi)$.
This implies that the MPML prior function is written as
\begin{equation}
\pi_M(\lambda,\psi)\,\propto\, \frac{1}{\lambda} \sqrt{ \frac{d^2}{d \psi^2} \log \Gamma(\psi) \,-\, \frac{1}{\psi} }\sqrt{\psi} \cdot \frac{\Gamma(n\psi)}{(n \psi)^{n\psi} } \exp(n\psi) \,=\, \pi_J(\lambda,\psi)\cdot\pi_m(\psi)
\label{eqn: MarPri}
\end{equation}
To evaluate the relation between the MPML and the reference prior functions we set 
$$
f(\psi,n) \,=\, \log \Gamma(n\psi) -(n-1/2)\psi\log(n \psi)\,+\,n\psi. $$
Then the reference prior function is written as $\pi_M(\lambda,\psi)\exp f(\psi,n)$.
Applying the asymptotic expansion formula of the logarithmic gamma function, we obtain $f(\psi,n) = (1/2)\log 2\pi \,+\,O((n\psi)^{-2})$.
Recall that this approximation is related to the Stirling formula.
This approximation is known to be accurate, even when $n\psi$ is not very large.
Thus we observe that the MPML and the reference prior functions are close to each other.

These examples indicate the close relationship between the MPML and the reference prior functions in the exponential dispersion model, though these priors were driven in largely different ways.
In fact, the MPML prior is designed for improving the conditional MLE, while the reference prior was introduced to define a non-informative prior in a rigorous sense.

\vspace*{3mm}
\noindent
{\bf Example 3.3} (Exponential family).
The exponential family of distributions contains a wide range of useful distributions.
It covers the exponential dispersion model in the previous example, and the canonical regression model reviewed briefly in the following example.

The density function having the sufficient statistic $(\bar t,\bar s)$ and the canonical parameter $(\xi,\psi)$ in the exponential family is written as
\begin{equation}
 p(\bx|\lambda,\psi) \,=\,\tprod \left[ \exp \{t_i \xi+ s_i \psi \,-\, M(\xi,\psi) \} a(x_i) \right]
\label{ExpFam}
\end{equation}
where $M(\xi,\psi)$ is a convex function and $\lambda=M_\xi(\xi,\psi)$ is the mean of $\bar t$.

This family satisfies various convenient properties such as the sufficient statistic and the complete statistic.
Studies on the dual structure of the parameter $(\lambda,\psi)$ were extensively explored by many authors including Hurzubazar (1956), Barndorff-Nielsen (1978) and Amari and Nagaoka (2007).
The importance of the dual structure becomes evident, and the structure was studied in depth.
The conditional likelihood inference based on the factorization property \eqref{FactAssump} owes to this dual structure.

This family provides us with textbook examples of the conditional MLE.
Since $\bar t$ is the MLE of $\lambda$ for an arbitrarily fixed $\psi$, the estimation orthogonality condition in \eqref{EstOrht} is satisfied.
The factorization property is also satisfied.
Thus a prior function $\pi_m(\psi)$ is defined.
When a prior function $\pi_m(\psi)$ is assumed, the posterior mode of $\psi$ is equal to the conditional MLE of $\psi$ given $\bar t$.
In addition, the posterior mode of $\lambda$ is given by $\hat \lambda_{Mod}=\bar t$.
These facts allow us to compare the posterior mode and the MLE of the parameter $(\lambda, \psi)$.
This simultaneous estimation of both the components of a parameter gives a broader view of the comparison study between the conditional MLE of $\psi$ and the MLE of $\psi$.
Such an attempt was found in Yanagimoto and Anraku (1989).

We suggested the default choice of the Jefferys prior as a non-informative prior and that of the canonical parameter as an estimand.
These choices imply that $\E_{\rm post}^N[(\xi,\psi)]$ - $(\hat\xi_{ML},\hat\psi_{ML})$ = $O(n^{-2})$, see Yanagimoto and Ohnishi (2021).
Coupling this relation with Theorem 2.1, we obtain the following asymptotic equivalence

\vspace*{3mm}
\noindent
{\bf Proposition 3.1}.

Suppose that the sampling density is in the exponential family \eqref{ExpFam}, and assume the MPML prior function \eqref{MRLDef}.
Under the condition that $\hat\psi_{CML} - \hat\psi_{ML}$ = $O(1/n)$ it holds that
\begin{equation}
\E_{\rm post}^M[(\xi,\psi)] \,-\, (\hat\xi_{ML},\hat\psi_{CML}) \,=\, O(n^{-2})
\label{Asymup2}
\end{equation}.

Next, we present inequalities between the posterior mean of the canonical parameter under the assumption of $\pi_M(\lambda,\psi)$ and the posterior mode under the assumption of $\pi_m(\psi)$.
For notational convention, we will write a predictor of the conditional density based on an observation $\bx$ as $pc(\by|t,\bx)$.
The component $\xi$ is written as $\xi(\lambda,\psi)$, when the parametrization $(\lambda,\psi)$ is employed.
We will identify an estimator and a plug-in predictor of the estimator to define the Kullback-Leibler divergence for notational simplicity. 

\vspace*{3mm}
\noindent
{\bf Proposition 3.2}.

The posterior mean $\E_{\rm post}^M [\KL\big( pc(\by|t,\bx), pc(\by|t,\psi) \big)]$  attains the maximum and the minimum at the plug-in predictor of the conditional MLE $ pc(\by|t,\hat\psi_{CML})$ and at the plug-in predictor of the posterior mean of $\psi$, $ pc(\by|t,\hat\psi)$, respectively.

\noindent
Proof. The conditional density $p(\by|t, \psi)$ is in the exponential family having the sufficient statistic $\bar s=\tsum s_i/n$ and the canonical parameter $\psi$.
The parallel treatments in the subsection 2.3 yields i).

\vspace*{3mm}
\noindent
{\bf Example 3.4} (Canonical link regression model).
Suppose that the $i$-th component of a sample vector $\bx=(x_1, \cdots x_n)$ follows a distribution in the exponential family with the canonical parameter $(c(\lambda_1), \cdots, c(\lambda_n) )$, and also that all the components distribute mutually independently.
Then the density function of $\bx$ is expressed as
\begin{equation}
\tprod \exp \{x_i c(\lambda_i) \,-\, M(c(\lambda_i))\} a(x_i). 
\label{ConReg1}
\end{equation}

Assume the simple canonical link regression model, $c(\lambda_i)=\alpha +\psi z_i$, where $z_i$'s are the exploratory variables satisfying the condition $\tsum(z_i- \bar z)^2\ne 0$.
Setting $t=\tsum x_i$ and $s=\tsum z_i x_i$, we can rewrite the model as
\begin{equation}
\exp \{t\alpha \,+\, s \psi \,-\,\tsum M(\alpha+z_i\psi) \}\tprod a(x_i). 
\label{ConReg2}
\end{equation}
We learn that this density is in the exponential family with the canonical parameter $(\alpha,\psi)$, and that the mean of $t$ is written as $\lambda=n\alpha+ \psi\tsum z_i $.
Thus the conditional MLE of $\psi$ given $t$ can be defined, and the corresponding MPML prior can also be driven.

An interesting point to be discussed arises, when the exploratory variables, $z_i$'s, are binary.
Suppose that $\lambda_i=\alpha + z_1 \psi$ for $i=1,\cdots, n_1$ and $\lambda_j=\alpha + z_2 \psi$ for $j=n_1+1,\cdots, n_1+n_2$ for $z_1\ne z_2$.
Then the two parametrizations, $(n_1\lambda_1+n_2\lambda_2, \psi)$ and $(\lambda_1,\lambda_2)$, are possible.
Consider two statistics, $t_1=\tsum x_i$ and $t_2=\tsum^{n_1} x_i/n_1$.
They are an ancillary statistic for $\psi$ and an s-ancillary statistic for $\alpha+\psi z_2$, respectively.
In this case the two models, \eqref{ConReg1} and \eqref{ConReg2}, are equivalent, and the two MLE's imply the equivalent plug-in predictors.

A familiar example of the canonical regression model with the binary explanatory variable is the two independent binomial models, $x \sim \Bi(n,p)$ and $y \sim \Bi(m,q)$.
Setting $t=x+y$, $s=x$, $\lambda=np+mq$ and $\psi=\log \{p(1-q)/q(1-p)\}$, we obtain an expression of the form in \eqref{ConReg2}.
This model can be straightforwardly extended to a logit regression model, which is a familiar regression model in analyzing the binary dataset.
Here we remark that the reference prior in these familiar models is not known.

\vspace*{5mm}
\noindent
{\bf 4. Extensions}
\setcounter{section}{4}
\setcounter{equation}{0}

A notable advantage of the present definition of a prior in \eqref{MRLDef} is in its possible straightforward extensions in various ways.
This fact makes the definition of the MPML prior function flexible.
We will find equivalent priors are driven from slightly different definitions.
A simple way is to apply the asymptotic normality of an ancillary statistic.
Other two are given by modifying the profile marginal density $pm(t|\lambda,\psi)$ and the Jeffreys prior.
Further, we add cases of multiple strata with a common $\psi$ and also of the marginal MLE.
We emphasize that the four extensions except for the firsr one are possible for a fixed sample size.
Possible various extensions are notable advantages of the present approach in contrast with the reference prior and also with the conditional MLE.

\vspace*{3mm}
\noindent
{ \bf 4.1. A prior based on asymptotic normality}.

First, we introduce a rule of thumb extension of the definition in \eqref{MRLDef} by applying the asymptotic normality.
An exact form of the MLE of an ancillary statistic $t$ and its marginal distribution are often complicated, but the asymptotic normality holds under weak regularity conditions.

Thus it is wise to apply the asymptotic normality to examine an approximated form of the MPML prior.
Necessary regularity conditions for the asymptotic normality of the MLE of the parameter will be assumed.
This conditions are satisfied in a wide class of familiar distributions.
Write the Fisher information matrix as $\I(\lambda, \psi)$.
The orthogonality condition on $(\lambda,\psi)$ yields the following asymptotic approximation
$$
{\hat\lambda_{ML} \choose \hat\psi_{ML} } \,\dot{\sim}\, \N \left( {\lambda \choose \psi}, \;\frac{1}{n} \begin{pmatrix} 1/\I_{11}(\lambda,\psi) & 0 \\ 0 & 1/\I_{22}(\lambda,\psi) \end{pmatrix} \right). $$

We set $t=\hat\lambda_{ML}$.
Since the asymptotic marginal density of $t$ follows $\N(\lambda,\I_{11}(\lambda,\psi)/n)$, $\pi_m(\psi)$ is approximated by $1/\sqrt{ \I_{11}(\hat\lambda_{ML},\psi) }$.
Choosing the Jeffreys prior function, we obtain that $\pi_M(\lambda, \psi)$ is approximated by $\sqrt{ \I_{11}(\lambda,\psi)\I_{22}(\lambda, \psi))/ \I_{11}(\hat\lambda_{ML},\psi) }$.

\vspace*{3mm}
\noindent
{\bf Proposition 4.1}.
Suppose that $\I_{11}(\lambda,\psi)$ is factored into 
$g_1(\lambda)g_2(\psi)$.
Then $\pi_m(\psi)$ is proportional to $g_2(\psi)$, and the asymptotic MPML prior function $\pi_{aM}(\lambda,\psi)$ is written as
\begin{equation}
\pi_{aM}(\lambda,\psi) \,=\, \sqrt{g_1(\lambda) \I_{22}(\lambda,\psi) }.
\label{AsymMRL}
\end{equation}
Note that the approximation does not necessarily require the factorization property for conditional inference.
This fact suggests possible generalizations to other models.
In contrast, it looks tough to extend the conditional MLE, when the factorization property is not satisfied.

Berger and Bernardo (1992) discussed a prior of the form $h(\lambda)\I_{22} $ for a suitably chosen function $h(\cdot)$ in their (1.3.3).
The present form in \eqref{AsymMRL} is an appealing choice.
Proposition 4.1 indicates that this asymptotically approximated prior can be regarded as an extension of the reference prior.
Note that explicit forms of the reference prior in the literature are driven in an asymptotic setting.

\vspace*{3mm}
\noindent
{\bf Examlpe 4.1}.
Examples of explicit forms of the reference prior were reviewed, and presented in a table in Garvan and Ghosh (1997).
It provides six examples of the reference prior.
Here we add the normal distribution to the table.
Interestingly, the asymptotic MPML prior in \eqref{AsymMRL} is equivalent to the reference prior in all of the seven examples.
Among them, we observed that the MPML prior is exactly equivalent to the reference prior in cases of the normal and the inverse Gaussian distributions, as in Examples 3.1 and 3.2.
The close relation was observed in the case of the gamma distribution, as also in Example 3.2.
A different interpretation on the relations between the MPML and the reference prior in the above cases will be given in the following subsection.

In a general model, the conditional likelihood given $\hat\lambda_{ML}$ depends on $\lambda$ as well as $\psi$.
Formal treatments of the Bayesian estimation under the MPML prior function were possible, as discussed above.
However, a formal extension of the asymptotic extension to the conditional MLE does not work well, since the components of the MLE $(\hat\lambda_{ML},\hat\psi_{ML})$ distribute asymptotically independently.

\vspace*{3mm}
\noindent
{ \bf 4.2. Choice of an ancillary statistic}.

We discuss here the alternative choice of an ancillary statistic instead of the MLE.
Recall that an ancillary statistic for the conditional MLE is not necessarily the MLE of $\lambda$. 
It is worth exploring another choice of a suitable ancillary statistic, which can make the definition of the MPML prior function simpler.
A reservation of this attempt concerns possible difference between the prior functions induced from different ancillary statistics.

A simple example is found in the case of an even sample size $n=2m$ in the Laplace distribution.
The uniqueness of the MLE of $\lambda$ does not hold in this case.
Write by $x_{(m)}$ the $m$-th order statistic in the ascending order.
Any point $\lambda$ in the interval $(x_{(m)},x_{(m+1)})$ maximizes the likelihood for every $\psi$.
It is possible to facilitate this inconvenience in various ways.
A method is to apply the limiting MLE in the family the location-power models $k_r(\psi)\exp \{\psi|x-\lambda|^r\}$ with a known index $r>1$.
The limiting MLE as $r$ tends to 1 is written as $t=(x_{(m)} +x_{(m+1)})/2$. 
This ancillary statistic looks appealing.
In fact, the limiting MLE is the usual definition of the sample median in the case of an even sample size.
It does not look necessary to justify this statistic $t$ as the MLE of $\lambda$.

To discuss suitable choices of an ancillary statistic, we discuss the case of the location-scale family.
This family includes the Laplace distribution, though necessary regularity conditions are not satisfied.
We suggest the choice of a sample median $\hat\lambda_{Med}$, which is equal to $x_{(m+1)}$ for $n=2m+1$ and to $\hat\lambda_{Med}= (x_{(m)}+x_{(m+1)})/2$ for $n=2m$, instead of the MLE of $\lambda$.
This choice will be examined in a general family in the following example.

\vspace*{3mm}
\noindent
{\bf Example 4.2} (Location-scale family).
Let $g(z)$ be a positive valued density function on the support $R$, and is assumed to be symmetric about 0.
Consider the location-scale family
\begin{equation}
 p(\bx|\lambda,\psi) \,=\,\tprod C \psi g\big( \psi (x_i \,-\, \lambda) \big) 
\label{LocScale}
\end{equation}
where $C$ is a normalizing constant.
The MLE of $\lambda$ is neither uniquely determined, nor its explicit form is often unavailable even when it is unique.
This fact results in a complicated form of the marginal likelihood.
A way to dissolve this problem is to apply the asymptotic normality of the MLE, as was discussed in the subsection 4.1. 

Another way is to examine the possible choice of the sample median $\tilde x_{Med}$ as an ancillary statistic for $\psi$ and an estimator of $\lambda$, $\hat\lambda_{Med}=\tilde x_{Med}$.
The sample median and the MLE of $\lambda$ are mostly different from each other, but their behaviors are expected to be close with each other in this model.
Intuitively, they are regarded as ancillary statistics for the scale component $\psi$.
Write the cumulative distribution function of $g(z)$ as $G(z)$.
Then that of $p(x|\lambda,\psi)$ is wirtten as  $P(x|\lambda,\psi)=G(\psi(x-\lambda))$.
For simplicity, we suppose that the sample size $n$ is odd, $2m+1$.
The marginal density of $\hat \lambda_{Med}$, $pm( \hat\lambda_{Med}| \lambda,\psi)$, is written as
\begin{equation}
\frac {\Gamma(n+1)}{\Gamma^2 (m+1)(m+1) }
G^m(\psi(\hat\lambda_{Med}-\lambda) )\psi g(\psi(\hat\lambda_{Med}-\lambda) ) (1-G(\psi(\hat\lambda_{Med}-\lambda) )^m 
\label{MargMed}
\end{equation}
This expression yields that $\pi_m(\psi) \,\propto \, 1/\psi $, since $pm(\hat\lambda_{Med}|\,\hat\lambda_{Med}, \psi) \propto \psi$, and $P(\hat\lambda_{Med}|\,\hat\lambda_{Med}, \psi)=G(0)$ for every $\psi$.
It follows that the elements of the Fisher information matrix are expressed as $\I_{11}\propto \psi^2$ and $\I_{22}\propto 1/\psi^2$, when formal definition of the matrix is available. 
This yields that the Jeffreys prior $\pi_J(\lambda,\psi)$ becomes the uniform prior.
Consequently, we obtain $\pi_M(\lambda,\psi)\propto 1/\psi$.
Note that the present treatments allow us to derive a prior without applying the asymptotic normality in the subsection 4.1.

When the sample size is even, $n=2m$, similar calculations yield $pm(\hat\lambda_{Med}|\,\hat\lambda_{Med}, \psi) \propto \psi$.
Since the Jeffreys prior for $(\lambda,\psi)$ is the uniform prior, as stated above, it follows that $\pi_M(\lambda,\psi)\propto 1/\psi$.

\vspace*{3mm}
\noindent
{ \bf 4.3. Choice of a non-informative prior}:

The Jeffreys prior was chosen as a default prior function matching the posterior mode under $\pi_m(\psi)$ and the posterior mean under the assumption of $\pi_M(\lambda,\psi)$.
This choice implies asymptotic equivalence up to the higher order $O(1/n^2)$ as in \eqref{Asymup2} when the sampling density is in the exponential family.
We observed that the MPML prior is equal to or asymptotically equivalent to the reference prior.

To explore the possible use of a non-informative prior other than the Jeffreys prior, we examine the use of the moment matching prior discussed in Ghosh and Liu (2011).
This class of prior functions is designed for attaining an accurate approximation between the MLE and the posterior mean under the moment matching prior.
They pointed out that the asymptotic approximation up to order $O(n^{-3/2})$ is realized by choosing a suitable estimand in various families of distributions.
Their aim is placed on justifying a non-informative prior by requiring the asymptotic equivalence of the posterior mean with the MLE.
No attempt was found in pursuing the superiority of the posterior mean under a moment matching prior over the MLE.

Corollary 2.1 can be applied to obtaining the asymptotic equivalence between the posterior mode and the posterior mean, even when the Jeffreys prior function is replaced by the moment matching prior function.

\vspace*{3mm}
\noindent
{\bf Example 4.3} (Exponential power distribution).
For a fixed value of $r (>1)$ we consider the density function in the general family of the exponential-power distributions indexed by $r$ as 
\begin{equation}
p_r(x|\,\lambda, \psi) \,=\,\frac{\psi^r} {2r \Gamma(r)} \exp\{-\psi |x\,-\,\lambda|^r \}
\label{LaplaceD}
\end{equation}
The value $r$ is regarded as an index of the heaviness of tail of a distribution in the family of distributions indexed by $r$.
When $r=2$, the subfamily indexed by $r$ is the normal family of distributions.
Setting $\xi=\psi^r$, we find that this family belongs to the location-scale family with the scale component $\xi$ in Example 4.2.
In addition, regularity conditions on the family are satisfied, and the results obtained in these two models can be applied.
On the other hand, the factorization property for the conditional inference \eqref{FactAssump} is not always satisfied.
In addition, the MLE can not be often written in an explicit form.

Two familiar distributions are driven as both the limits of at 1 and $\infty$, which are the Laplace and the uniform distributions, respectively.
They do not necessarily the regularity conditions, but satisfy convenient properties.

The density function of the Laplace distribution is written as $p_1(x|\,\lambda, \psi) =(\psi/2)\exp\{-\psi |x\,-\,\lambda| \}$.
The MLE of $\lambda$ is uniquely determined, and is the same as the sample median $\hat\lambda_{MLE}=\hat \lambda_{Med}$ when the sample size $n$ is odd.
When the sample size is even $n=2m$, the MLE is not formally defined, but a reasonable, unique definition is possible.
A definition is possible by taking the limit of the sequence of the MLE's in the sampling density $p_r(\bx|\lambda, \psi)$ at $r=1$.
This definition yields that $\hat\lambda_{ML}=(x_{(m)}+x_{(m+1)})/2$, which is the the sample median.
This fact allows us to employ the sample median as an ancillary statistic for $\psi$, which implies that 
$\pi_m(\psi)\propto 1/\psi$. 
Note that $\lambda$ is estimation orthogonal to $\psi$.
However, the factorization property does not hold, and the conditional MLE of $\psi$ does not make sense.

A problem to derive the MPML prior is to yield the Jeffreys prior.
This problem is dissolved by the usual convention; the definition of the Fisher information matrix is not defined by the Hesse matrix of minus of the logarithmic likelihood but by the correlation matrix of the likelihood estimating function.
Another way to facilitate this problem is to define the information matrix by a limit of the Jeffrey prior of a sequence of density functions, as above.
Consequently, the Jeffreys prior function becomes proportional to a constant, and it follows that $\pi_M(\lambda,\psi) \propto 1/\psi$.

The other limit at $r=\infty$ of the exponential power distribution is the uniform distribution on the support $(\lambda-1/\psi,\, \lambda+1/\psi)$, $\U(\lambda-1/\psi,\, \lambda+1/\psi)$.
Write the smallest and the largest order statistics as $x_{(1)}$ and $x_{(n)}$, respectively.
Then $(x_{(1)},x_{(n)})$ is a sufficient statistic for $(\lambda,\psi)$.
The MLE of $\lambda$ is written as the mid-range $(x_{(1)}+x_{(n)})/2$ for every $\psi$.
Thus $\lambda$ is estimation orthogonal to $\psi$.
However, this family does not satisfy the factorization property.
The uniform distribution can formally be regarded to be in the location-scale family in \eqref{LocScale}, though the support of the density depends on the unknown  parameters.

Formal treatments as above imply that $\pi_m(\psi)$ and $\pi_M(\lambda,\psi)$ are both proportional to $1/\psi$.
This prior function is equivalent to the formal limit of the reference prior function in the sequence of the exponential-power distributions as $r$ tends to $\infty$.
Let an estimand be $(\psi\lambda,\psi)$, which is equivalent to $(\lambda,\psi)$.
Then it follows that.
\begin{equation}
\hat\psi \,=\, \frac{2(n-1)} {(n+1)(x_{(n)}-x_{(1)}) }.
\label{EstUnif}
\end{equation}
As stated above, the posterior mean of $\lambda$ becomes the midrange. 
Both estimators are written as functions of sufficient statistics $x_{(1)}$ and $x_{(n)}$.
Further, the estimators $\hat\lambda$ and $1/\hat\psi$ are unbiased, and are uniformly minimum variance unbiased estimators (UMVUE).
In contrast, the MLE of $1/\psi$ is $(x_{(n)}-x_{(1)})/2$, which is downward biased.

An interesting observation is that the induced non-informative prior function from three different ways is commonly proportional to $1/\psi$ for every $r$; one is the reference prior, the second is the asymptotic MPML prior in the subsection 4.1 and the third is a prior based on the choice of the median as an ancillary prior in the previous subsection.

\vspace*{5mm}
\noindent
{\bf 4.4. A common $\psi$ through strata}

Next, we discuss briefly an extension to a practical model, where the conditional MLE is expected to perform favorably.
The necessity of the separate inference is evident, when the multiple strata having a common component $\psi$ through strata and stratum-wise components $\lambda_k$ for $k=1,\cdots,K$.
Let $\bx_k$ be a subsample of size $n_k$ from the $k$-th stratum with the density function $p(\bx_k|\lambda_k,\psi)$.
Note that such a situation appears in various fields of applications such as the paired comparison.
Consider the situation that $K$ tends to infinity and $n_k$ is bounded.
Then the MLE of $\psi$ is likely to be inconsistent (Neyman-Scott, 1948). However, the conditional MLE is consistent, when the factorization property for each density $p(\bx_k|\lambda_k,\psi)$ holds.

The MPML prior can be straightforwardly extended, and is expressed as
$$
\pi_M(\lambda_1,\cdots,\lambda_K,\psi)\,\propto\, \tprod \left\{ \frac{1} 
{ pm \big(t_k | \hat\lambda_{kML}, \psi \big) }
\pi_J(\lambda_k,\psi) \right\}. $$

\vspace*{5mm}
\noindent
{\bf 4.5. Marginal MLE}

Another technique of separate inference similar to the conditional MLE is the marginal MLE.
The following factorization property, instead of \eqref{FactAssump}, is assumed on the sampling density,  
\begin{equation}
p(\bx|\lambda,\psi) \,=\, pc(\bx|t,\lambda,\psi)pm(t|\psi)
\label{MargFacA}
\end{equation}
for every $\bx$ and $\theta$.

Parallel treatments are possible.
A prior function can be defined in terms of the reciprocal of $pc(t|\hat\lambda_{ML}(t,\psi),\psi)$ as $\pi_{c}(\psi) \,\propto\, 1/pc(\bx|t,\hat\lambda_{ML},\psi)$. 
A prior function corresponding to the MPML prior is expressed as $
\pi_{C}(\lambda,\psi) \,\propto\, \pi_{c}(\psi) \pi_J(\lambda, \psi) $.
Note that the use of the Jeffreys prior is justified by the asymptotic relation corresponding to Theorem 2.1, when the regularity condition on the posterior mode $\theta_r = \Argmax \pi_c(\psi)p(\bx|\lambda,\psi)$ is satisfied.

Simple examples of the marginal MLE are found in the exponential dispersion model in Example 3.2.
Set $t=n N(\bar x)  - \sum N(x_i)$ and $s=\bar x$.
Then $t$ and $s$ distribute independently, and the conditional MLE of $\psi$ is regarded also as the marginal MLE.
Another example is found in the von Mises distribution (Shou, 1978).

\vspace*{5mm}
\noindent
{\bf 5. Discussions.}
\setcounter{section}{5}
\setcounter{equation}{0}

Our primary goal is to improve the conditional MLE by introducing a pair of novel priors.
Amazingly, the MPML prior is close to the reference prior in familiar examples.
Since the conditional MLE and the reference prior are very important concepts in the current estimation theory, we give comments and critiques on them, based on the present studies.
These efforts are expected to elucidate the role of the present approach.

\vspace*{3mm}
\noindent
{\bf 5.1. Conditional MLE}.

The conditional MLE is based on the conditional likelihood given an ancillary statistic, when the factorization property \eqref{FactAssump} is satisfied.
The original definition of an ancillary statistic, which is referred to as an s-ancillary statistic, is that the conditional likelihood of $\bx$ given the statistic does not depend on the parameter, see Cox and Reid (1987) and Lindsey (1996) for example.
Cases where the conditional likelihood depends only on $\psi$ are widely employed.
When the conditional MLE is expected to perform more favorably than the MLE, the PML prior $\pi_m(\lambda,\psi)$ is not the uniform prior.
Then the MPML prior is different from the Jeffreys prior.
The posterior mean of a suitably chosen estimand under the MPML prior is expected to perform better than that under the Jeffreys prior. 

We summarize here three points relating the MPML prior.
One concerns the notion of a nuisance parameter.
In the exponential family discussed in Example 3.2 a nuisance component appears as a component of the mean (or expectation) parameter.
In contrast, the component of interest appears as a canonical parameter.
As a special case of the exponential dispersion model, a nuisance component denotes the mean of observation, while the component of interest denotes the dispersion.
So far as our experience in practical applications, our primary interest is more likely to be in the mean component than in the dispersion component.
Another problem in this concern pertains to the custom that a nuisance component is not estimated.
It was our own experiences that many researchers strongly raised this assertion.
However, there does not give rise to any inconvenience by estimating all the component of a parameter.
In contrast, advantages of estimating all the components of the parameter are obvious.
One is that an estimator implies a plug-in predictor.
Recall that a predictor plays an important role in the recent theory of statistics.

The second point pertains to the amount of information contained in an ancillary statistic.
When the marginal density of $t$ depends on $\psi$.
It is reasonably believed that the ancillary statistic actually contains some amount of information about $\psi$.
The problem is how to extract the information.
It is believed also that the superiority of the conditional MLE over the MLE of $\psi$ is realized by discarding the marginal likelihood.
The introduction of the prior $\pi_M(\lambda,\psi)$ can be regarded as an attempt to extract some amount of information from the marginal likelihood of an ancillary statistic.

The last point concerns the role of the conditional MLE.
The conditional MLE was accepted due to facilitating the unexpected behavior of the MLE, as pointed out clearly by Neyman and Scott (1948).
Thus researchers' interest focused on the superiority of $\hat\psi_{CML}$ over $\hat\psi_{ML}$.
Least attentions have been paid on the improvement of the conditional MLE.
The introduction of the MPML prior is an attempt to improve the conditional MLE under Bayesian setting.

\vspace*{3mm}
\noindent
{\bf 5.2. Reference prior}

The reference prior looks persuasive among candidates of a non-informative or an objective prior.
It has been widely accepted as a reasonable non-informative prior, when it exists.
It does not require any restrictive conditions such as the factorization property in \eqref{FactAssump}.
Instead, it is necessary to fix which of the components is of ineterst, or to determine both the components are equally of interest.

A rigorous definition of the reference prior is given in Berger, Bernardo and Sun (2009).
Their definition of the reference prior is based on the maximization of the Kullback-Leibler divergence $\KL (\pi(\psi|\bx),\pi(\psi))$.
This divergence is useful in theoretical statistics, which is employed in the optimality property in the subsection 2.3.
In the literature, see Robert (2001) for example, the reference prior is favored because it is right Haar measure invariant, while the Jeffreys prior is left Haar measure invariant.
They are fully conceptual, and no direct relation with a possibly favorable performance of an induced estimator is claimed.

There are two critiques against this prior; one is its limited applicability, and the other is the lack of an affirmative claim regarding expected advantages induced from the assumption of the reference prior.
A non-informative, objective prior may be hoped to satisfy such strong requirements.
It becomes common to evaluate the asymptotic behavior to derive explicit forms of the reference prior, see Ghosh (2011) for example.

Conceptually, the notion of a non-informative prior emphasizes that it does not contain any information on the parameter in a model.
The definition of the reference prior is intuitively appealing, but it is difficult to discuss whether it is ideal or not.

The present approach is more practical.
We attempt to explore a prior yielding a favorable performance of the posterior mean in the frequentist context.
The performance varies with a chosen loss.
This fact requires us to choose a suitable estimand.
Note that the reference prior for $(\lambda,\psi)$ is invariant under the component-wise transformation $(g_1(\lambda), g_2(\psi))$ for smooth and strictly monotone transformations.
Thus it is necessary to choose a suitable estimand so that the induced estimator performs favorably. 

It is our understanding that the reason why the reference prior is widely accepted is not necessarily its persuasive definition but due to its favorable performance of the induced procedures.
We noted in Example 3.2 that when the sampling density is in the exponential dispersion model the performance of the posterior mean under the assumption of the reference prior is close to the posterior mode under the assumption of the prior $\pi_m(\lambda,\psi)$.
Since the posterior mode of $\psi$ is the same as the conditional MLE of $\psi$ given $t=\bar x$, the posterior mean under the assumption of the reference prior is expected to perform better than that under the Jeffreys prior.
This fact motivated us to attempt to introduce a novel prior.

\vspace*{3mm}
\noindent
{\bf References}
\begin{list}{}
 {\leftmargin .28in \rightmargin .15in \itemindent -.25in \itemsep 0.0in \parsep .0in}
\vspace*{-1mm}
\item
Amari, S., Nagaoka, H., 2007.  
Methods of Information Geometry. Am. Math. Soc., Rhode Island
\item  
Barndorff-Nielsen, O.E., 1978.
 Information and Exponential Families in Statistical Theory. John Wiley and Sons, New York.  
\item
Berger, J.O., Bernardo J.M., 1992.
On the development of the reference prior method (with discussion). Bayesian Statistics 4 (J.M. Bernardo, J.O. Berger, A.P. Dawid, A.F.M. Smith eds).  Oxford University Press, 35-60.

\item
Berger, J.O., Bernardo J.M., Sun, D., 2009.
The formal definition of reference priors. Ann. Statist. 37 (2), 905-938.
\item
Bernardo, J.M., 1979.
Reference posterior distributions for Bayesian inference.
J. R. Stat. Soc.. Ser. B, 41 (2), 113-147. 
\item
Corcuera, J.M,, Giummole, F., 1999. A generalized Bayes rule for prediction. Scand. J. Statist. 26 (2), 265-279.
\item 
Cox, D.R., Hinkley, D.V., 1974. 
Theoretical Statistics.  Chapman and Hall, London.
\item  
Cox, D., Reid, N., 1987. Parameter orthogonality and approximate
conditional inference (with discussion). J. Roy. Statist. Soc. Ser. B 49 (1), 1-39.
\item
Garvan, C.W., Ghosh, M., 1997. Noninformative priors for dispersion models.  Biometrika 84 (4), 976-982.
\item
Ghosh, M., 2011. Objective priors: An introduction for frequentists.
Statist. Sci., 26 (2), 187-202.

\item
Ghosh, M., Liu R., 2011. Moment matching priors. Sankhya Ser. A 73 (2), 185-201
\item 
Huzurbazar, V.S., 1956. Sufficient statistics and orthogonal parameters. Sankhya 17 (3), 217-220.
\item
Jorgensen, Bent., 1997. The Theory of Dispersion Models. Chapman \& Hall, London.
\item  
Kullback, S., 1959. Information Theory and Statistics., 
Wiley, New York. 
\item
Lindsey, J.K., 1996.
Parametric Statistical Inference.  Clarendon Press, Oxford.
\item
Miyata, Y., 2004. Fully exponential Laplace approximations using
asymptotic modes. J. Am. Statist. Assoc. 99 (468), 1037-1049,
\item  
Neyman, J., Scott, E.L., 1948.  Consistent estimates based on
partially consistent observations. Econometrica 16, 1-32.
\item
Robert, C.P., 2001.
The Bayesian Choice Second ed. Springer, New York.
\item 
Yanagimoto, T., Anraku, K., 1989, Possible superiority of the conditional MLE over the unconditional MLE. Ann. Inst. Statist. Math. 41 (2), 269-278. 
\item
Yanagimoto, T., Ohnishi, T., 2009. 
Bayesian prediction of a density function in terms of $e$-mixture. J. Statist. Plann. Inf. 139 (9), 3064-3075. 
\item
Yanagimoto, T., Ohnishi, T., 2011. 
Saddlepoint condition on a predictor to reconfirm the need for the assumption of a prior distribution. 
J. Statist. Plann. Inf. 41 (5), 1990-2000. 
\item
Yanagimoto, T., Ohnishi, T., 2021.
A characterization of Jeffreys' prior with its implications to likelihood inference.
Pioneering Works on Distribution Theory: In Honor of Masaaki Sibuya, ( N. Hoshino et al. eds), Springer, 103-121.

\end{list}

\end{document}